\documentstyle[psfig]{mn}

\newif\ifAMStwofonts

\ifoldfss
  \ifCUPmtlplainloaded \else
    \NewTextAlphabet{textbfit} {cmbxti10} {}
    \NewTextAlphabet{textbfss} {cmssbx10} {}
    \NewMathAlphabet{mathbfit} {cmbxti10} {} 
    \NewMathAlphabet{mathbfss} {cmssbx10} {} 
  \fi
  \ifAMStwofonts
    \ifCUPmtlplainloaded \else
      \NewSymbolFont{upmath} {eurm10}
      \NewSymbolFont{AMSa} {msam10}
      \NewMathSymbol{\upi}     {0}{upmath}{19}
      \NewMathSymbol{\umu}     {0}{upmath}{16}
      \NewMathSymbol{\upartial}{0}{upmath}{40}
      \NewMathSymbol{\leqslant}{3}{AMSa}{36}
      \NewMathSymbol{\geqslant}{3}{AMSa}{3E}

      \let\leq=\leqslant 
       
    \fi
  \fi
\fi 

\ifnfssone
  \newmathalphabet{\mathit}
  \addtoversion{normal}{\mathit}{cmr}{m}{it}
  \addtoversion{bold}{\mathit}{cmr}{bx}{it}
  \newmathalphabet{\mathbfit} 
  \addtoversion{normal}{\mathbfit}{cmr}{bx}{it}
  \addtoversion{bold}{\mathbfit}{cmr}{bx}{it}
  \newmathalphabet{\mathbfss} 
  \addtoversion{normal}{\mathbfss}{cmss}{bx}{n}
  \addtoversion{bold}{\mathbfss}{cmss}{bx}{n}
  \ifAMStwofonts
    \ifCUPmtlplainloaded \else
      \UseAMStwoboldmath
      \makeatletter
      \new@mathgroup\upmath@group
      \define@mathgroup\mv@normal\upmath@group{eur}{m}{n}
      \define@mathgroup\mv@bold\upmath@group{eur}{b}{n}
      \edef\UPM{\hexnumber\upmath@group}
      \new@mathgroup\amsa@group
      \define@mathgroup\mv@normal\amsa@group{msa}{m}{n}
      \define@mathgroup\mv@bold\amsa@group{msa}{m}{n}
      \edef\AMSa{\hexnumber\amsa@group}
      \makeatother
      \mathchardef\upi="0\UPM19
      \mathchardef\umu="0\UPM16
      \mathchardef\upartial="0\UPM40
      \mathchardef\leqslant="3\AMSa36
      \mathchardef\geqslant="3\AMSa3E

      \let\leq=\leqslant 

    \fi
  \fi
\fi 

\ifnfsstwo
  \DeclareMathAlphabet{\mathbfit}{OT1}{cmr}{bx}{it}
  \SetMathAlphabet\mathbfit{bold}{OT1}{cmr}{bx}{it}
  \DeclareMathAlphabet{\mathbfss}{OT1}{cmss}{bx}{n}
  \SetMathAlphabet\mathbfss{bold}{OT1}{cmss}{bx}{n}
  \ifAMStwofonts
    \ifCUPmtlplainloaded \else
      \DeclareSymbolFont{UPM}{U}{eur}{m}{n}
      \SetSymbolFont{UPM}{bold}{U}{eur}{b}{n}
      \DeclareSymbolFont{AMSa}{U}{msa}{m}{n}
      \DeclareMathSymbol{\upi}{0}{UPM}{"19}
      \DeclareMathSymbol{\umu}{0}{UPM}{"16}
      \DeclareMathSymbol{\upartial}{0}{UPM}{"40}
      \DeclareMathSymbol{\leqslant}{3}{AMSa}{"36}
      \DeclareMathSymbol{\geqslant}{3}{AMSa}{"3E}

      \let\leq=\leqslant 

    \fi
  \fi
\fi 

\ifCUPmtlplainloaded \else
  \ifAMStwofonts \else 
    \def\upi{\pi}
    \def\umu{\mu}
    \def\upartial{\partial}
  \fi
\fi

\def\ltsima{$\; \buildrel < \over \sim \;$}
\def\gtsima{$\; \buildrel > \over \sim \;$}

\title[A second turn-off in NGC 1868]
{On the nature of a secondary main sequence turn-off in the rich LMC cluster NGC 1868}
\author[B. Santiago et al.]
  { B.~Santiago,$^1$ L. Kerber,$^1$ R. Castro,$^1$ R. de Grijs$^2$ \\
$^1$Universidade Federal do Rio Grande do Sul, Instituto de F\'\i sica, 91501-970 Porto Alegre, RS Brasil\\
  $^2$Institute of Astronomy, Madingley Rd.,Cambridge CB3 0HA, UK}

\begin{document}
\maketitle
\label{firstpage}
\begin{abstract}  
Evidence for a second main-sequence turn-off in a deep colour-magnitude 
diagram of NGC 1868 is presented. The data were obtained with HST/WFPC2 and
reach down to $m_{555} \simeq 25$. Besides the usual $\tau \simeq 0.8~Gyr$ 
turn-off found in previous analyses, another possible turn-off 
is seen at $m_{555}
\simeq 21$ ($M_{V} \simeq 2.5$), which is consistent with an age of
$\tau \simeq 3~Gyrs$. This CMD feature stands out clearly especially when 
contaminating field LMC stars are statistically removed. 
The background subtracted CMD also 
visibly displays a red subgiant branch extending about
1.5 mag below the younger turn-off and the clump of red giants.
The significance of the secondary turn-off in NGC 1868 was confirmed with
Monte-Carlo simulations and bootstrapping techniques. 
Star-counts in selected regions in the cluster CMD indicate
a mass ratio of old population/young population in the range $5\%$
\ltsima $M_{old} / M_{young}$ \ltsima $12\%$, depending on the mass function
slope. The existence of such a subpopulation in NGC 1868 is significant even
in the presence of uncertainties in background subtraction.
The possibility that the secondary turn-off is associated with the field star 
population was examined by searching for similar features in CMDs of field
stars. Statistically significant excesses of stars redwards of the 
main-sequence were found in all such fields in the range $20$ \ltsima 
$m_{555}$ \ltsima $22$. These however are much broader features that do
not resemble the main-sequence termination of a single population.
We also discuss other alternative explanations for the feature at $m_{555}
\simeq 21$, such as unresolved binarism, peculiar stars or CMD discontinuities
associated with the B\"ohm-Vitense gap.

\end{abstract}


\section{Introduction}

NGC 1868 is a rich Large Magellanic Cloud (LMC) cluster located 
at $\alpha = 5^h~14^m$ and
$\delta = -63^o~57'$ (J2000), approximately 6$^\circ$ away from the LMC's 
centre. Its age has been estimated both from ground-based photometry and
spectroscopy, often yielding discrepant results covering the
$3~10^8~yrs$ \ltsima $\tau$ \ltsima $10^9~yrs$ range (Flower et al. 1980,
Hodge 1983, Bica \& Alloin 1986, Chiosi et al. 1986, Olszewski
et al. 1991, Corsi et al. 1994 and references therein). 
Most age estimates come from the interpretation of optical
colour-magnitude diagrams (CMDs), usually based on the positions 
of either the main-sequence turn-off (MSTO) or the clump of red giants 
(RC) or both.
However, none of the CMDs available until now have been deep enough to
allow probing the main sequence at magnitudes as faint as $V \simeq 22$ 
with small photometric errors (see the web page on 
www.ast.cam.ac.uk/STELLARPOPS/LMCdatabase
for a detailed list of references on NGC 1868 and other rich LMC clusters
in the HST Cycle 7 program GO7307). 

In this paper we investigate a possible second, fainter and therefore
older, MSTO at $V \simeq 21$, based on a deep CMD of NGC 1868 built from
HST/WFPC2 data.
This second population of stars may be the result of a strong interaction
with another cluster, or of 
the capture of a lower-mass, older cluster by NGC 1868. In fact,
there is evidence for a growing number of clusters in the LMC which 
may have undergone strong interactions or mergers (Kontizas et al. 1993, 
Dieball \& Grebel 1998, Leon et al. 1999, Dieball, Grebel \& Theis 2000).
Confirmation of a merger event in a cluster's history, however, requires
detailed photometric and/or spectroscopic data.
Sagar et al (1991), based on a CMD of
NGC 2214 with two apparent supergiant branches, suggested that it was made
up of two distinct populations.
However, Lee (1992), Bhatia \& Piotto (1994) and Banks et al (1995),
based on larger and more accurate photometry, found a single population in
the CMD of NGC 2214.
Given that signatures of mergers or strong interactions remain 
observable in clusters for at least 1 Gyr (de Oliveira, Bica \& Dottori 2000), 
NGC 1868, if confirmed as a merger product,
may be a good laboratory for studying 
the dynamical effects and the final products of such events. Hence, it is 
essential to investigate closely the possibility that it is made up of
two distinct populations.

Other explanations for a feature similar to a MSTO exist. Unresolved 
binaries, for instance, are usually brighter and redder than the main 
sequence of single stars. Therefore, an enhanced fraction of unresolved 
binaries could result 
in such a feature. Sudden changes in stellar structure, such as the onset
of convective envelopes, may lead to CMD features such
as the B\"ohm-Vitense gap, that may also mimic a MSTO (B\"ohm-Vitense 1970).
These possibilities are also considered in the analysis of our deep
NGC 1868 CMD, as well as of other similar CMDs.

The paper outline is as follows: in \S 2 we describe the HST/WFPC2 data used,
both on NGC 1868 and on control field areas. We also discuss
the process of removing contamination by field stars from the on-cluster CMD. 
In \S 3 we test the statistical significance of the candidate
secondary MSTO by means
of bootstrapping realizations. In \S 4 we use isochrones and star counts in 
selected CMD areas to extract useful information about this presumed secondary
population in NGC 1868.
In \S 5 we explore the possibility that the candidate MSTO is related to
field stars, making use of CMDs obtained in control field areas. We also
explore the alternative mechanisms that could yield to features similar
to a MSTO at that position.
Our main conclusions are presented in \S 6.

\begin{figure*}
\begin{center}
\centerline{\psfig{file=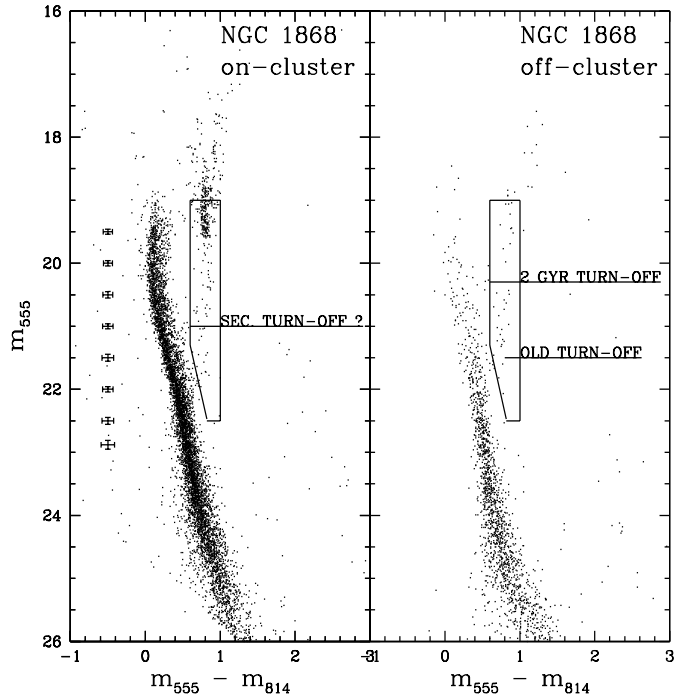,height=10cm,width=9.5cm,angle=0}}
\end{center}
\caption{ On-cluster (left panel) and off-cluster (right panel) colour-magnitude diagrams. The large trapezium box indicates the locus of evolved stars, whose field star subtraction was carried out separately from that of MS stars. Empirically determined photometric uncertainties are shown on the left panel.}
\end{figure*}

\begin{figure*}
\begin{center}
\centerline{\psfig{file=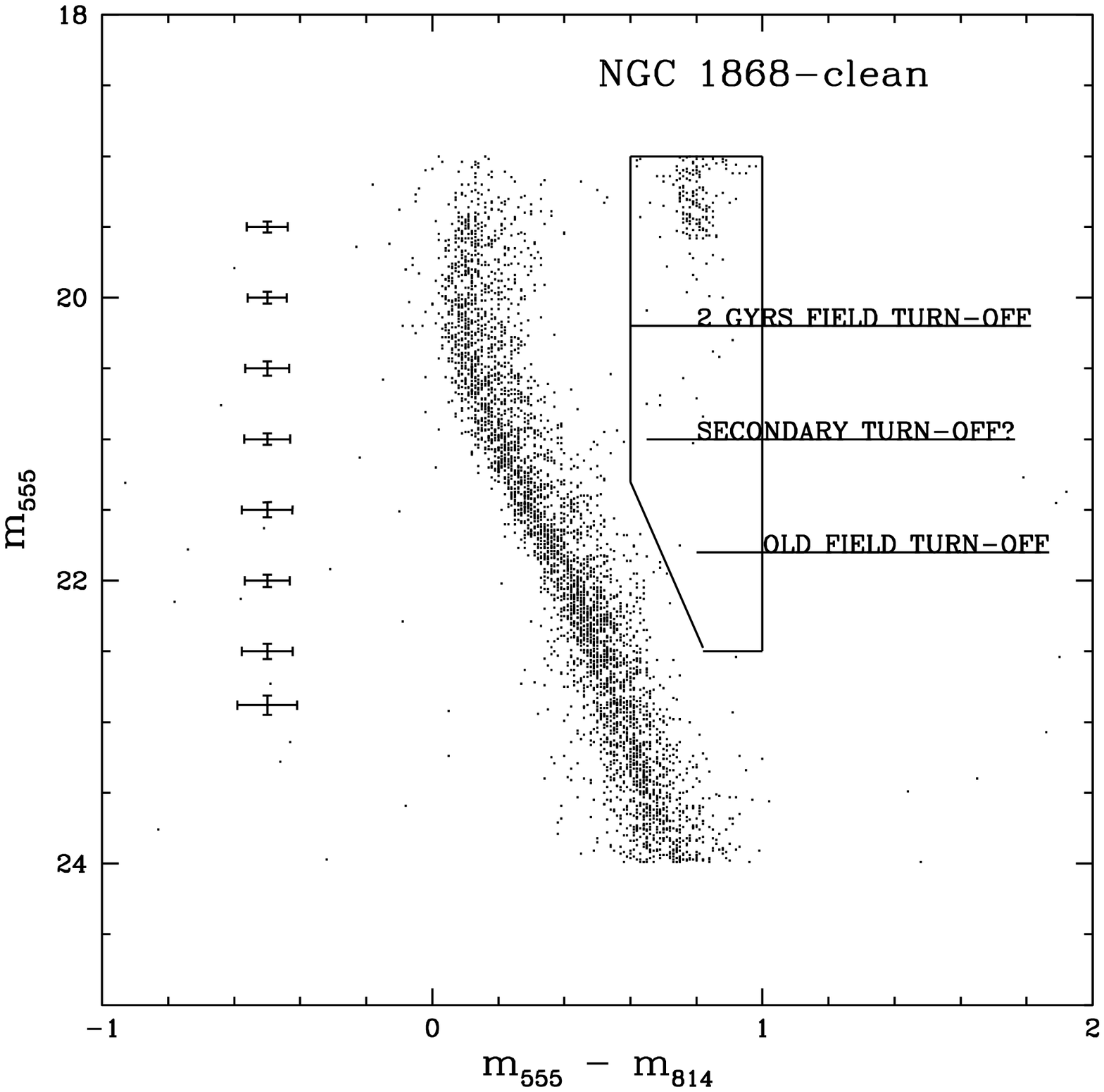,height=10cm,width=9.5cm,angle=0}}
\end{center}
\caption{Colour-magnitude diagram resulting from subtracting field stars from the on cluster data. }
\end{figure*}

\section[]{The data}

The WFPC2 on-cluster data are described in more detail in 
Santiago et al. (2001).
The off-cluster data used for field star subtraction were reduced and 
analyzed by Castro et al. (2001). In both cases the images were combined
and calibrated using the standard pipeline procedure (Holtzman et al. 
1995a,b). Photometry was carried out using DAOPHOT tasks and is
described in detail by Santiago et al (2001) and Castro et al (2001). 
Zero-points,
CTE and aperture corrections were also applied, again following standard
procedures (Santiago et al 2001, de Grijs et al 2002a,b). 

Figure~1 shows the resulting CMDs. The left panel shows stars belonging
to the on-cluster images. Two such images were taken, 
one with the Planetary Camera (PC)
at the centre of NGC 1868 and the other at its half-light radius. They
are the CEN and HALF fields as defined by Santiago et al. (2001). The CMD shown
represents the final sample, including stars from both HALF and CEN fields 
and with no repeats. The total on-cluster solid angle is 
approximately 7.5 arcmin.
As explained in Santiago et al. (2001), the
stars in the region common to the HALF and CEN fields had
two independent photometric measurements and were thus used 
to determine uncertainties as a function of
$m_{555}$ magnitudes and $m_{555} - m_{814}$ colour. A total of 731
stars were found in the overlap region between the HALF and CEN fields.
These empirical error
determinations are an essential part of our upcoming analysis.
The uncertainties were computed in bins 0.5 mag wide in both $m_{555}$ and
$m_{814}$. In each magnitude bin we 
computed the mean value and standard deviation ($\sigma$) of
the distribution of magnitude differences. The typical uncertainty of a
single measurement was then assumed to be $\sigma / \sqrt{2}$.
The photometric uncertainty analysis is described in more detail 
in Kerber et al (2002).

The right panel in Figure 1 shows stars in the off-cluster area, 
located 7.3' away
from the centre of NGC 1868 and previously studied by
Castro et al (2001). This off-cluster field corresponds
to a single WFPC2 field, covering about 5 arcmin.
Both CMDs include only objects that were classified
as stellar sources in the classification schemes presented 
in Santiago et al (2001) and Castro et al (2001). No
additional cleaning of remaining spurious objects or background galaxies
was made, resulting in some objects 
located well away from the cluster main-sequence (MS), the red giant 
branch (RGB) and the RC. 
Most of them should be contaminating unresolved 
background galaxies or faint stars in our Galaxy. 

The on-cluster stars display
a clear MS that terminates at $m_{555} \simeq 19.2$. Previous works find
the MS termination at $V = 19.2-19.3$ (Corsi et al 1994, Brocato et al
2001). This is also 
roughly the magnitude of the RC ($m_{555} - m_{814} \simeq 0.8$), which 
contains He burning stars and is largely 
dominated by the cluster stars. In both
panels, a clear RGB stretches downwards from the RC position
to the bottom of the subgiant branch (SGB) at $m_{555} \simeq 22$,
$m_{555} - m_{814} \simeq 0.8$. These stars mostly
belong to the old
($\tau$ \gtsima $10~Gyrs$) LMC field star population (Holtzman et al. 1997, 
Castro et al. 2001, Smecker-Hane et al 2002). 
Besides the main cluster and old field turn-offs, there 
are two additional features that may be interpreted as
MSTOs in the CMD: one has mostly
field stars with $m_{555} \simeq 20.3$ and had already been identified
by Castro et al. (2001) as a $\tau \simeq 2~Gyrs$ field population; 
the other is fainter, at $m_{555} \simeq 21$, $0.4 < m_{555} - m_{814} < 0.5$
and consists entirely of stars in the on-cluster 
data. It is this latest feature that we concentrate on, as a candidate
NGC 1868 subpopulation. The MSTOs described here are indicated in Figure~1.

\begin{figure*}
\begin{center}
\centerline{\psfig{file=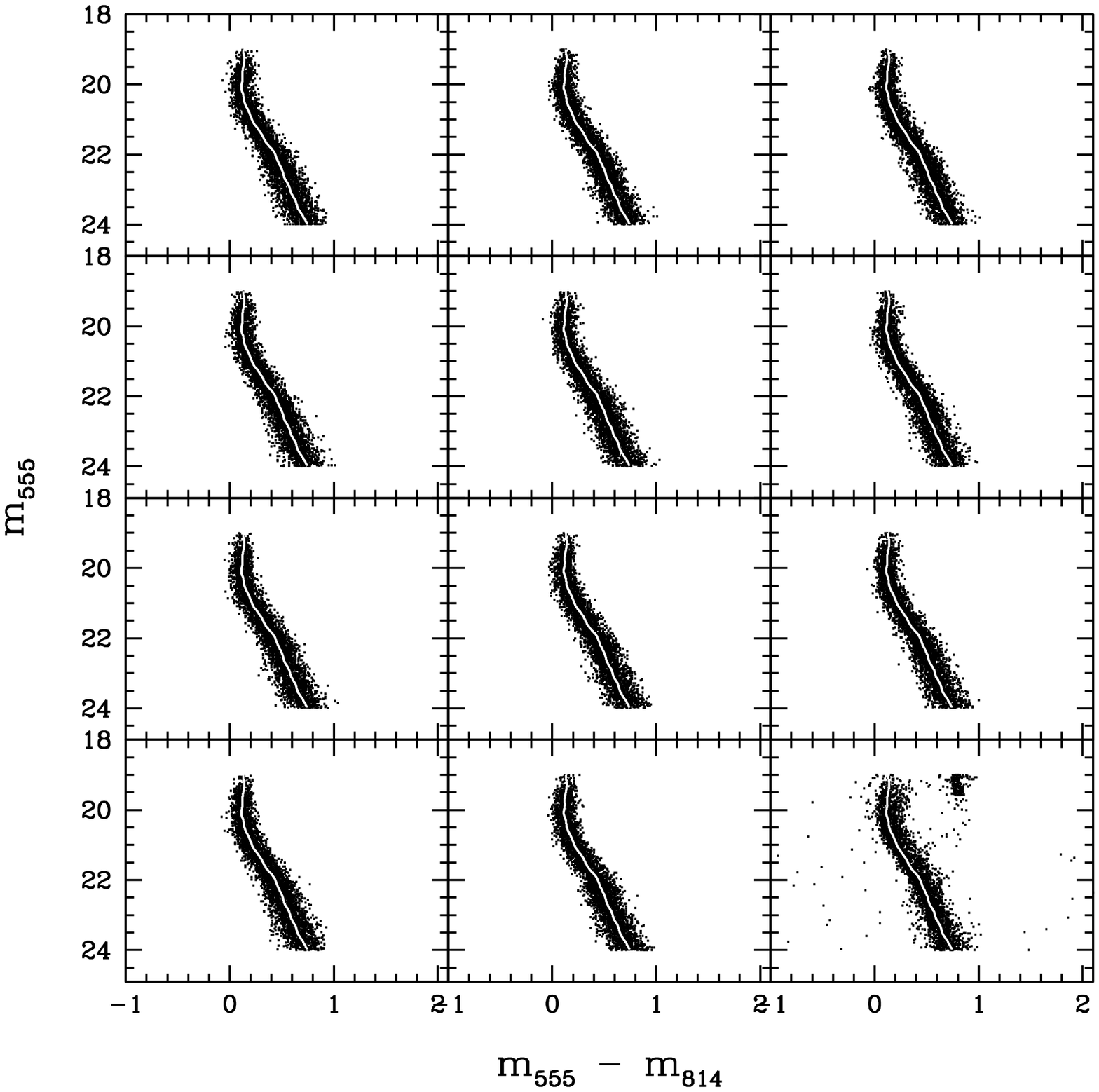,height=20cm,width=19.0cm,angle=0}}
\end{center}
\caption{A sample of artificial CMDs built from the NGC 1868 MS fiducial line (white line) and error distribution. The lower-right panel shows the actual NGC 1868 data, for comparison}
\end{figure*}

\subsection[]{Removing contaminating field LMC stars}

In order to clean the left panel of Figure~1 from background 
field contamination, we statistically remove
the control off-cluster CMD from the on-cluster one. 
Several procedures for doing so have been considered. 
One method is to match the
off-cluster stars to corresponding stars in the on-cluster 
image using the estimated probabilities that two stars could be
independent photometric measurements of each other. 
In order to estimate these 
matching probabilities, we use the empirically determined 
photometric uncertainties in both images and assume a 
Gaussian error distribution. Cluster stars are then randomly 
removed according to their
probability of matching any of the off-cluster stars. In this
approach a reliable estimate of the photometric uncertainties is 
very important.
As pointed out previously, we used the independent magnitude 
measurements for the
stars in common between the CEN and HALF fields to empirically estimate 
standard deviations ($\sigma$)
in the error distribution at different magnitude bins
(Santiago et al 2001, Kerber et al 2002).

An alternative way of removing contaminating field stars is to bin both
on-cluster and off-cluster data in magnitude 
(or in magnitude and colour) and to subtract the histogram of the latter 
from that of the former.
At each bin, a subset of the on-cluster stars,
corresponding to the subtracted histogram,
is randomly selected. The two approaches have been tested and yield
similar results.
As our main concern is to assess the existence of a second, older population
superposed to the dominant NGC 1868 population, the branch of evolved stars is
critical, since it provides the best opportunity to single out stars from each
individual population. Therefore, field subtraction was carried out 
separately for the MS and for the SGB/RGB locus. This latter is marked 
on Figure~1. 

A more detailed discussion about the issue of field stars removal
is presented in an upcoming paper (Kerber et al. 2002).

The cleaned CMD is shown in Figure~2. The CMD is now cut-off at the faint end
($m_{555} > 24$) as this region is not relevant to the current analysis. 
Below this limit, sampling incompleteness would only further complicate field
star subtraction. We
also cut it at bright magnitudes ($m_{555} < 19$) to avoid
regions where saturation effects start to take place ($m_{555} \simeq 19$
and $m_555 \simeq 17.8$ for the HALF and CEN fields, respectively).
The statistical removal of field stars has depleted 
the subgiant branch in the range $21$ \ltsima $m_{555}$ \ltsima $22$.
A residual number
of SGB stars brighter than $m_{555} \simeq 21$ but much fainter than
the cluster RC remains. As this excess of SGB stars
reltive to the field is located at brighter magnitudes than
the candidate second MSTO, it is a first evidence for the reality of this
feature.
As for the secondary MSTO itself, at $m_{555} \simeq 21$, it was
left untouched, which reflects
the absence of field stars close to it. 
This again supports the reality of the second population in NGC 1868.

\section[]{Statistical significance of the candidate secondary MSTO}

In this section we will assess whether the feature we tentatively
identify as a secondary MSTO in the CMD of NGC 1868 is statistically 
significant.
More specifically we address the question of whether a similar feature could
originate from a single main sequence through a random 
realization of photometric errors in the data.
We explore this possibility by means of Monte-Carlo realizations of the cleaned
NGC 1868 CMD. In each realization we collapse the cluster MS onto a fiducial
line and redistribute the data, using the measured photometric 
uncertainties and
assuming Gaussian error distributions in magnitude and colour.

The main sequence fiducial line is determined by taking 
the median $m_{555} - m_{814}$ colour at different magnitude bins. We 
apply a 3$\sigma$ clipping to the colour distribution in order to eliminate
outliers and iterate until convergence. We initially define this
MS line in bins spaced by 0.2 mag. A cubic spline
interpolation is then used to fill this line in a 
much narrower ($\Delta m_{555} = 0.02$) binning. 
Each true MS star is then randomly assigned to one of the points along the
MS fiducial line, using a two-dimensional Gaussian probability 
function whose standard deviations are
given by the star's measured magnitude and colour uncertainties. 
We then redistribute the data back onto the CMD plane,
the magnitude and colour of each artificial star again being a random 
realization of 
the same Gaussian distribution of errors.

One hundred such Monte-Carlo simulations were carried out. A subset with
11 of the resulting
CMDs is shown in Figure~3. The actual data are shown in the lower-right panel
for comparison. The MS fiducial line is shown as a white line for guidance.
The artificial CMDs have similar width as the data but are unable
to reproduce the apparent turn-off at $m_{555} \sim 21$, or the residual
of the old field population turn-off ($m_{555} \sim 22$), whose stars are
farther to the red of the MS and much more clumped together than any 
set of points in any of the simulated CMDs. The simulations also fail
to reproduce the main cluster MSTO at $m_{555} \simeq 19$ and
$m_{555} - m_{814} \simeq 0.3$. This proves that {\it scatter due to 
photometric errors alone cannot account for these features seen in the data}.
Notice that these experiments do not explicitly account for
the effect of unresolved binaries on the CMD. 
The fact that the distribution of true stars is
skewed redwards from the MS fiducial line is likely caused by unresolved
binaries. Incorporation of unresolved binaries would help spread out the
simulated stars towards redder colours, but is unlikely to reproduce
the features mentioned above, especially if we consider that part of the
effect is already incorporated in the position of the MS fiducial line.
We discuss the effect of binaries in more detail in \S 5.2. 

\begin{figure*}
\begin{center}
\centerline{\psfig{file=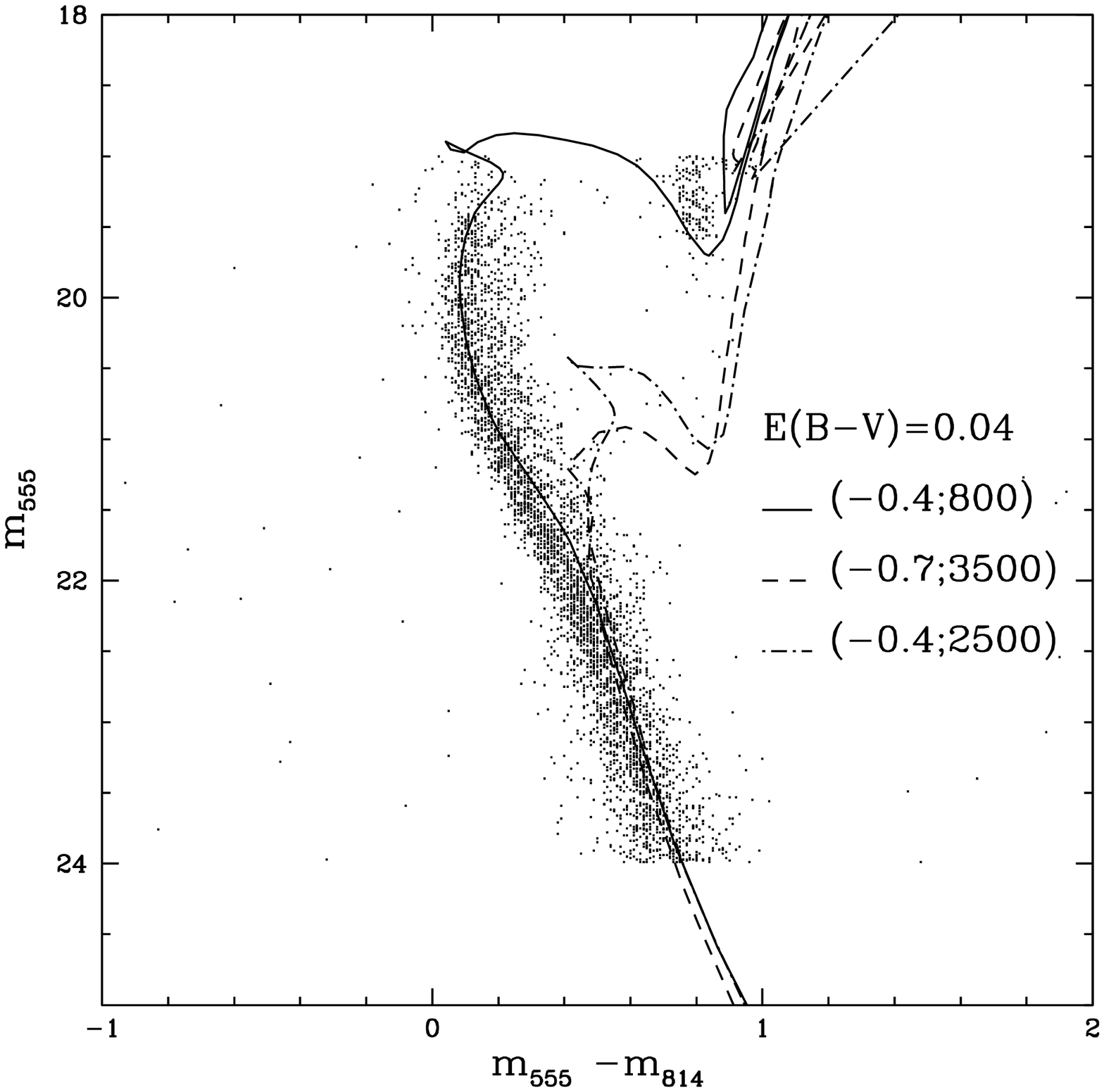,height=10cm,width=9.5cm,angle=0}}
\end{center}
\caption{Field subtracted cluster colour-magnitude diagram. Some Padova isochrones are superimposed to the data; their [Fe/H] and age (in Myrs) are indicated in the figure, as well as the adopted reddening value.}
\end{figure*}

\section[]{The candidate NGC 1868 second population}

Assuming that the second MSTO at $m_{555} \sim 21$ is real and belongs
to the cluster, one
wishes to determine what are the main characteristics of this 
population in NGC 1868.

In Figure~4 we again show the cleaned NGC 1868 CMD. 
Superimposed to the data we show three Padova isochrones
(Girardi et al. 2000).
In plotting the isochrones, we assume a distance modulus of
$m - M = 18.5$ to the LMC (Panagia et al. 1991). The data were extinction
corrected assuming E(B-V) =0.04 as indicated. Conversion from
E(B-V) to E(555-814) was done as described by Holtzman et al (1995a,b).
The $[Fe/H]$ and age (in Myrs) of each isochrone are shown in the figure. 
The [-0.4,800] isochrone fits the main NGC 1868 population.
The other isochrones are attempts to fit the
second turn-off and SGB.
The younger and more metal-rich isochrone with [Fe/H] = -0.4 and
an age of 2500 Myrs provides the best fit
to both turn-off and SGB regions.

\begin{figure*}
\begin{center}
\centerline{\psfig{file=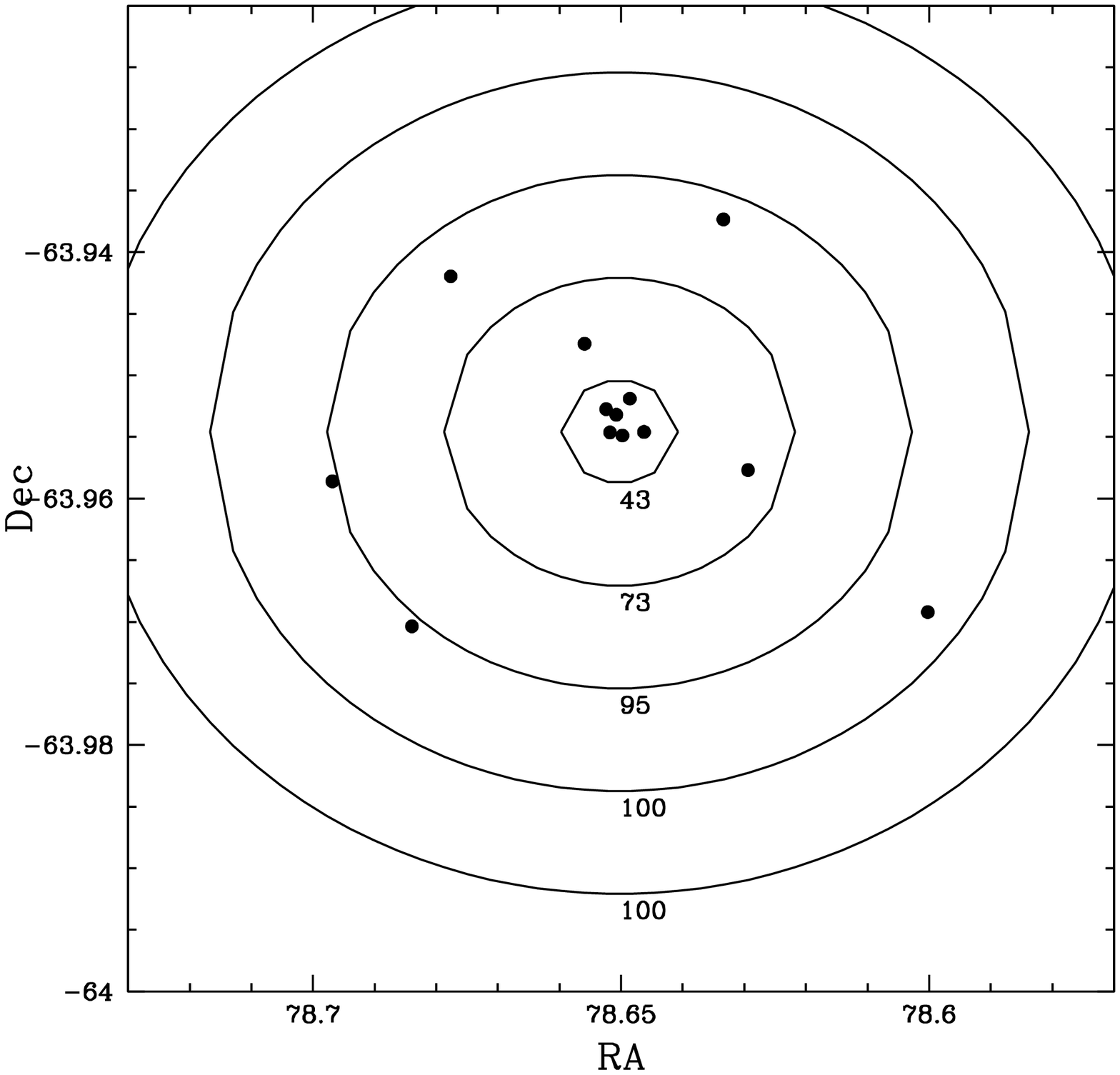,height=10cm,width=9.5cm,angle=0}}
\end{center}
\caption{Dots: projected distribution of stars located at the CMD position of the secondary MSTO. The contours are circles centred on the cluster. The
numbers shown vertically are the percentage of all cluster stars interior to each circle.}
\end{figure*}

Using star counts and this best fitting isochrones,
we can estimate the
mass fraction of this candidate older NGC 1868 population relative to the 
dominant and younger one.
There are 11 stars in the range $2.5$ \gtsima 
$M_{555}$ \gtsima $1.5$ ($21.0$ \gtsima $m_{555}$ \gtsima $20.0$) 
and located along the SGB. Besides these, 13 stars are
located within the CMD area whose limits are: $20.85 \leq m_{555}
\leq 21.3$, $m_{555} - m_{814} \leq 0.6$ and $m_{555} \leq 5.333~(m_{555} -
m_{814}) + 18.55$. This corresponds to the region 
where the second MSTO is clearly detached from the MS.
Assuming that all such $N_{old} = 24$ stars are in fact associated
with an older population belonging to NGC 1868 and
using the best fitting isochrone, we infer the masses
that correspond to the limiting range in absolute magnitudes: 
they are $m_{min} \simeq 1.36~m_{\odot}$ for the basis of
MSTO position and $m_{max} \simeq 1.49~m_{\odot}$ for $M_{555} = 1.5$
($m_{555} = 20$) along the SBG.  
As for the dominant cluster population we find $N_{young}$ = 732 stars  
along the cluster MS with $M_{555} < 2$ ($m_{555} < 20.5$)
and fainter than the MS termination (we cut it at $m_{555} = 19$). These
limits correspond to $1.63$ \ltsima $m/m_{\odot}$ \ltsima $2.10$ for 
the best fitting
isochrone to the NGC 1868 upper MS ([Fe/H],age(Myrs) = -0.4,800). 
Assuming the cluster's 
present day mass function (PDMF) to be a power-law
with fixed slope $\alpha$ within the mass ranges considered, 
the mass ratio will be given by

$${ {M_{old}} \over {M_{young}} } = { {N_{old}} \over {[ m_{max,old}^{1-\alpha} - 
m_{min,old}^{1-\alpha} ]} }~{ {[ m_{max,young}^{1-\alpha} - 
m_{min,young}^{1-\alpha} ]} \over {N_{young}} }$$

Assuming a Salpeter value for the MF ($\alpha = 2.35$), and the limiting
mass values quoted above for each of the subpopulation, we then infer 
$M_{old} / M_{young} = 0.06$. For $\alpha = 1.5$ we have 
$M_{old} / M_{young} = 0.08$ and
for $\alpha = 3.2$, $M_{old} / M_{young} = 0.05$. The mass ratios are also
fairly insensitive to adopting the ([Fe/H],age(Myrs)) = (-0.7,3500) Padova
isochrone for the secondary population; in this case, the
old population mass range would be $1.22 < m/m_{\odot} < 1.29$ and
$M_{old} / M_{young}$ would change by no more than 50\% for any
choice of $\alpha$. The mass ratio estimates are summarized in Table~1.

\begin{figure*}
\begin{center}
\centerline{\psfig{file=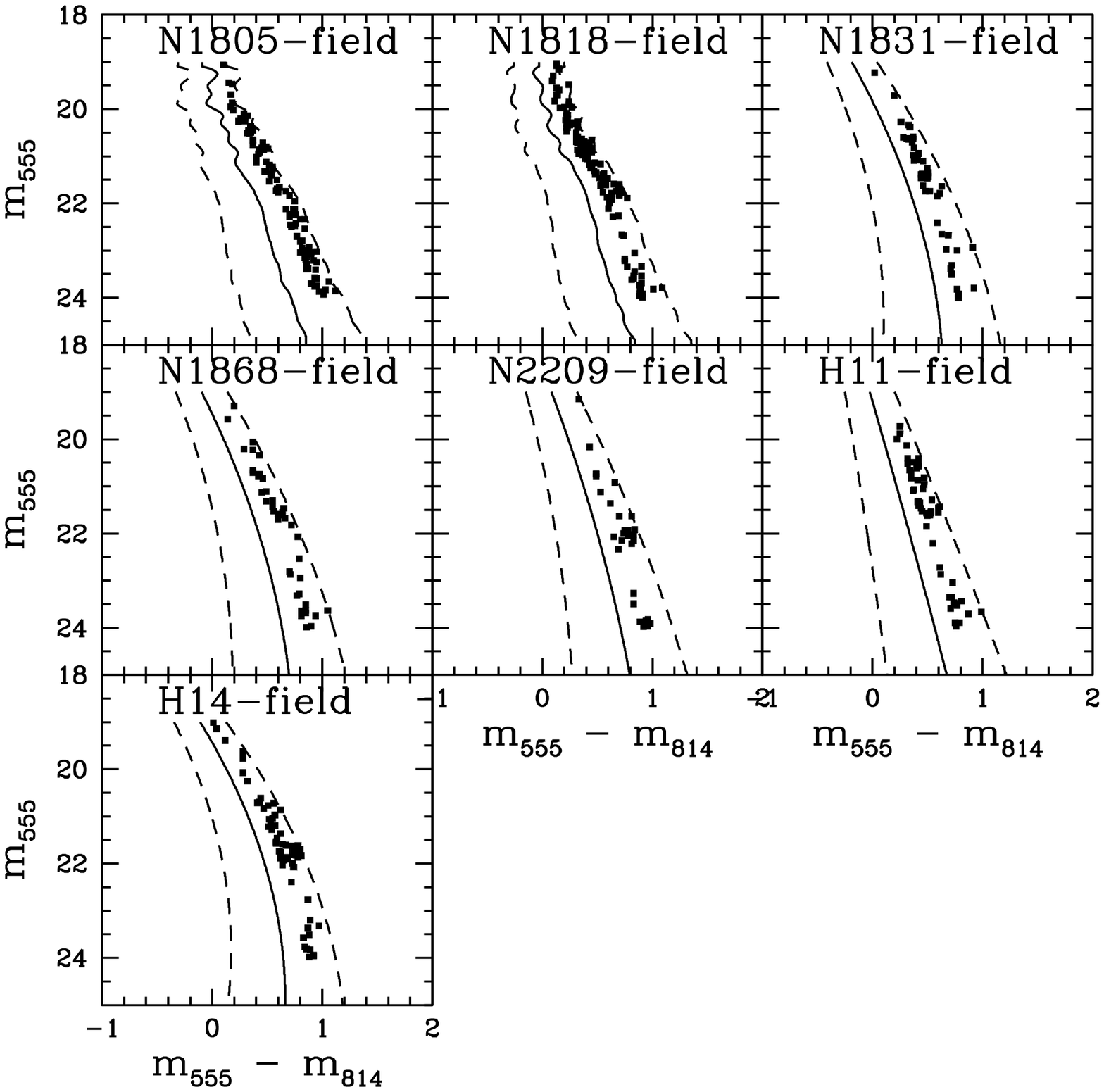,height=20cm,width=19cm,angle=0}}
\end{center}
\caption{CMDs of field LMC stars with the 5\% highest values of $n_{sig}$ as defined in the text. Each panel corresponds to one particular field in the LMC as indicated. The central solid line is the MS fiducial line and the dashed lines correspond to its $\pm~5\sigma$ deviations. }
\end{figure*}

In addition to the PDMF slope and the model mass-luminosity relation, 
another source of uncertainty in the mass ratio quoted
above is the residual contamination by field stars to the $N_{old}$ 
counts in the SGB region. 
Previous experiments with the field star subtraction (see \S 2.1)
methods have led to
variations of up to $35\%$ in the resulting value of $N_{old}$.
If we assume this to be the uncertainty caused by field contamination,
a similar relative error is expected to propagate into 
$M_{old} / M_{young}$.
Notice that this is a conservative reasoning, since residual contamination by
field LMC stars should also affect the younger population.
We should also point out that this $35\%$ uncertainty is larger than the 
Poisson fluctuation expected for the $N_{old}$ counts.

{\bf
\begin{table}
\caption{Mass ratios for the two subpopulations of NGC 1868 stars for different choices of PDMF slope and assumptions for the age, metallicity and mass range of evolved stars in the secondary population}
\begin{tabular}{c c c c c c c c}
\hline
isochrone & SGB mass range (solar) & $\alpha$ & $M_{old} / M_{young}$ \\
\hline
$[-0.4,2500]$ & 1.36/1.49 & 2.35 & 0.06 \\
$[-0.4,2500]$ & 1.36/1.49 & 1.50 & 0.08 \\
$[-0.4,2500]$ & 1.36/1.49 & 3.20 & 0.05 \\
$[-0.7,3500]$ & 1.22/1.29 & 2.35 & 0.09 \\
$[-0.7,3500]$ & 1.22/1.29 & 1.50 & 0.12 \\
$[-0.7,3500]$ & 1.22/1.29 & 3.20 & 0.06 \\
\hline
\end{tabular}
\end{table} 
}

\begin{figure*}
\begin{center}
\centerline{\psfig{file=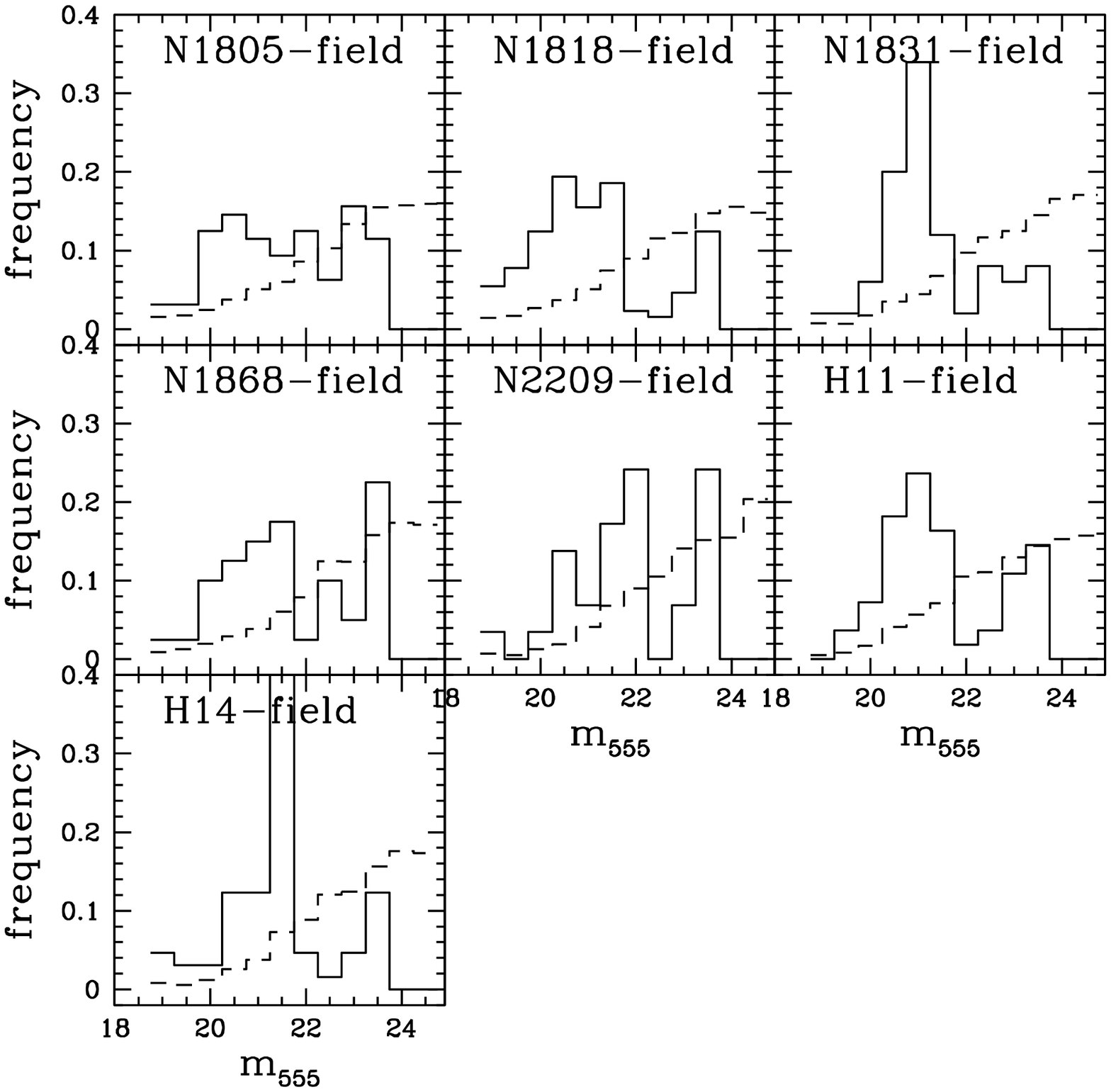,height=20cm,width=19cm,angle=0}}
\end{center}
\caption{Frequency distribution of the stars shown in the previous figure as a function of $m_{555}$ magnitude. Again each panel corresponds to one of the LMC fields, as indicated. The histogram shown as a dotted line corresponds to the frequency distribution of all MS stars}
\end{figure*}

An important question is how the stars in this candidate secondary 
population
are distributed throughout the cluster. Differences in spatial distribution
between the two populations could indicate a recent merger, in which the
two systems have not yet attained dynamical equilibrium.
Figure~5 shows the projected distribution of the stars located
at the second turn-off; they are the same 13 stars
used in estimating the $M_{old}/M_{young}$ ratio earlier in this section.
The 11 more evolved stars along the SBG were left out of Figure~5
because they represent just a statistical excess relative to field counts;
hence, they are not necessarily true members of the older population.
The contours represent circles around the cluster centre and the attached
labels show the percentage of all cluster stars located inside them. 
We note that, within
the fluctuations expected by the small numbers, the projected distribution
of these stars is consistent with that of the entire cluster.

\section {Alternative interpretations}

We have so far been interpreting the CMD feature at $m_{555} \simeq 21$
and $m_{555} - m_{814} \simeq 0.4$ as a second MSTO associated with 
NGC 1868.
In this section we discuss other possible interpretations.

\subsection {Is it a population associated with the LMC field?}

One obvious alternative is that the feature in the NGC 1868 CMD
is in fact a MSTO, but
associated with stars belonging to the general LMC field population rather than
to the cluster itself. Notice however that this is unlikely, since most
contaminating field stars have been statistically removed from the on-cluster
CMD. Furthermore, Figure 5 shows that the stars located at the position of the
second turn-off are concentrated towards the cluster centre, a strong evidence 
in favour of a cluster origin for them.
On the other hand, the star formation history in the LMC is known
to be complex, especially in the last few Gyrs, when the bar was formed
(Gallagher et al 1996, Elson et al 1997, Smecker-Hane et al 2002).
Thus, small scale variations in the field CMDs are possible, making
field subtraction more uncertain.

One way to investigate possible MSTOs in field populations is to search for
similar features in the 7 CMDs of field stars
studied by Castro et al (2001). These are
fields located 7.3' from the target clusters of the GO7307 HST project.
In fact, Castro et al have visually identified possible
turn-offs, with ages in the range $2$ \ltsima $\tau$ \ltsima $4$ Gyrs, 
in several of these fields.
In addition to those, turn-offs associated with an old ($> 10$ Gyrs) 
population were found in
all field CMDs. In some cases, visual inspection of the CMDs revealed
a broadening in the MS for $m_{555}$ \ltsima $21.5$, 
indicative of continuous star formation in the LMC.

In order to quantify these visual impressions we tried to identify
features in the field CMDs containing stars that are highly detached
from the MS line, similar to the presumed NGC 1868 MSTO.
One obvious
such locus is the SGB/RGB region, which is always present in the CMDs
of field LMC stars. It thus has to be eliminated from
the analysis {\it a priori}. 
We must also avoid contamination from background galaxies and faint 
stars belonging to the Galaxy. Unresolved galaxies should be limited
to faint magnitudes, $m_{555}$ \gtsima $22.5$, as brighter ones are 
visibly extended in our WFPC2 images. From counts of faint compact
galaxies, one expects $\sim 50$ contaminating galaxies 
within the $22.5 \leq m_{555} \leq 25$ range in our on-cluster field 
(Abraham et al 1996). As for foreground stars, a similar
number is expected in the entire observed CMD (Santiago et al 1996). 
Together both types of
contaminating sources contribute with a few percent of the total sample 
and will be spread out in the CMD.
 
We therefore proceeded as follows. For each field a fiducial line representing 
the CMD MS was defined
in the same way as described in \S 3. 
A low-order polynomial was fit to the MS fiducial line of most
fields in order to smooth out the wiggles caused by 
noise in the median $m_{555} - m_{814}$ value at each
$m_{555}$ bin. 
For NGC 1805 and NGC 1818, which have much larger
numbers of MS stars than the other fields, 
especially at the bright end ($20.5 < m_{555} < 19$), the raw MS line
was used. Evolved stars, possible background galaxies and faint stars 
belonging to the Galaxy were then
eliminated by cutting out all CMD objects beyond $\pm 5~\sigma$ from
the MS line, where $\sigma$ is the empirically determined standard deviation 
in the 
$m_{555} - m_{814}$ colour distribution at each magnitude.
For the remaining stars, we computed the number of standard deviations
by which each star is detached from the MS fiducial line, $n_{sig} = 
\Delta (m_{555} - m_{814}) / \sigma$.

In Figure~6, we show the stars whose $n_{sig}$ values fall
at the 95\% position or beyond in each LMC field. 
The number of stars in each panel 
varies from 30 to 130.
These stars would be $> 2~\sigma$ (and $< 5~\sigma$)
events of a Gaussian error distribution in colour.
The MS fiducial line and the $\pm 5~\sigma$ lines are also shown for guidance.
Assuming that the photometric uncertainties have been adequately measured
over the entire $m_{555}$ range and that these stars just reflect
the high tail of the error distribution, we would expect them to follow the
distribution of MS stars.
Therefore, the shape of the distribution of such stars may
reveal features that are not accounted for by errors alone.
As an example, a clumped distribution of such stars at some magnitude 
range, as compared to 
the smooth distribution of MS stars, would indicate the existence
of features such as a MSTO.

\begin{figure*}
\begin{center}
\centerline{\psfig{file=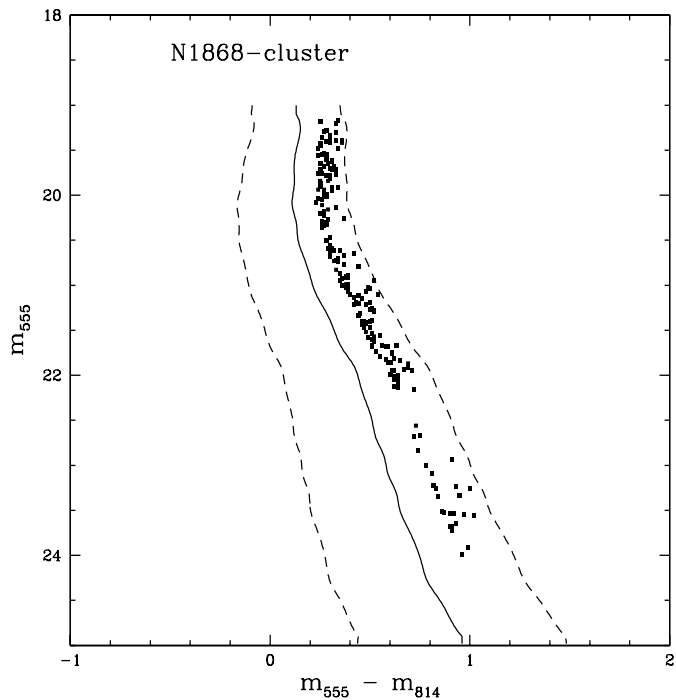,height=10cm,width=9.5cm,angle=0}}
\end{center}
\caption{CMDs of the NGC 1868 stars with the 5\% highest values of $n_{sig}$ as defined in the text. The central solid line is the MS fiducial line and the dashed lines correspond to its $\pm~5\sigma$ deviations. }
\end{figure*}

\begin{figure*}
\begin{center}
\centerline{\psfig{file=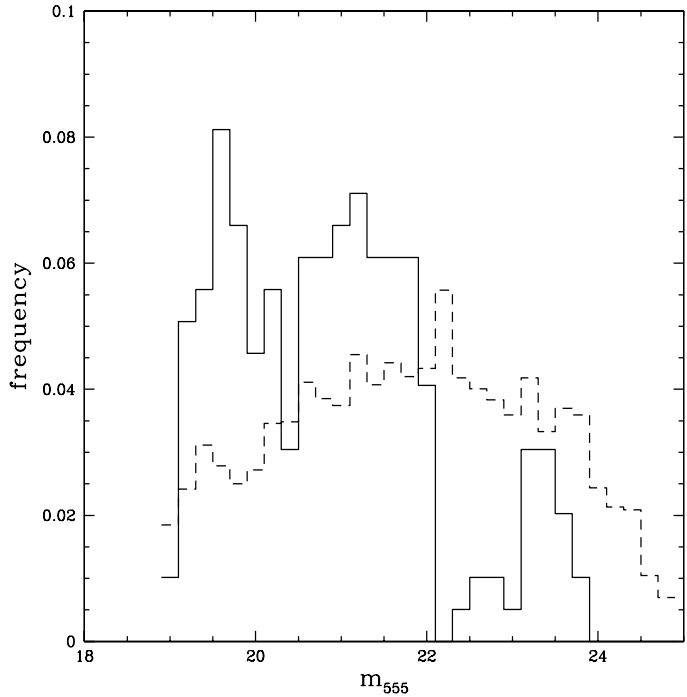,height=10cm,width=9.5cm,angle=0}}
\end{center}
\caption{Solid line: Frequency distribution of the stars shown in the previous figure as a function of $m_{555}$ magnitude. The dotted line corresponds to the frequency distribution of all MS stars in NGC 1868}
\end{figure*}

\begin{figure*}
\begin{center}
\centerline{\psfig{file=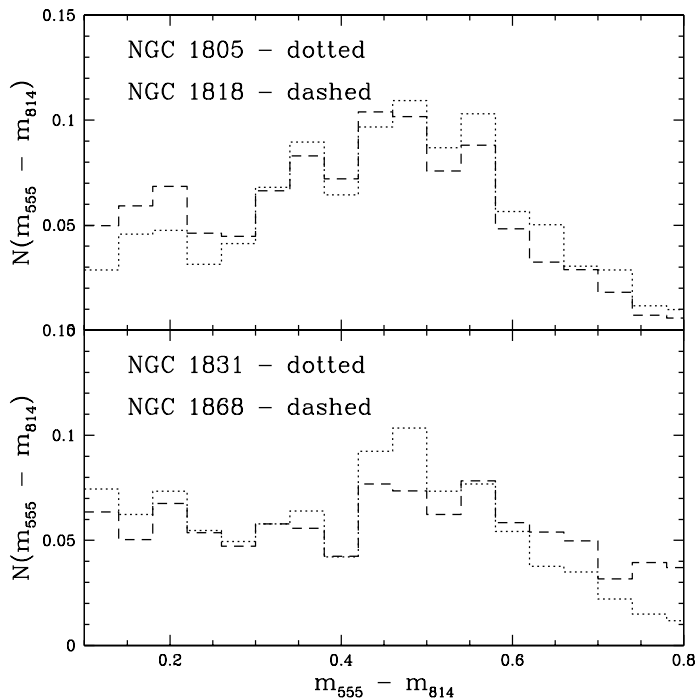,height=10cm,width=9.5cm,angle=0}}
\end{center}
\caption{Colour distributions for NGC 1805, NGC 1818 (upper panel), NGC
1831 and NGC 1868 (lower panel). }
\end{figure*}

\begin{figure*}
\begin{center}
\centerline{\psfig{file=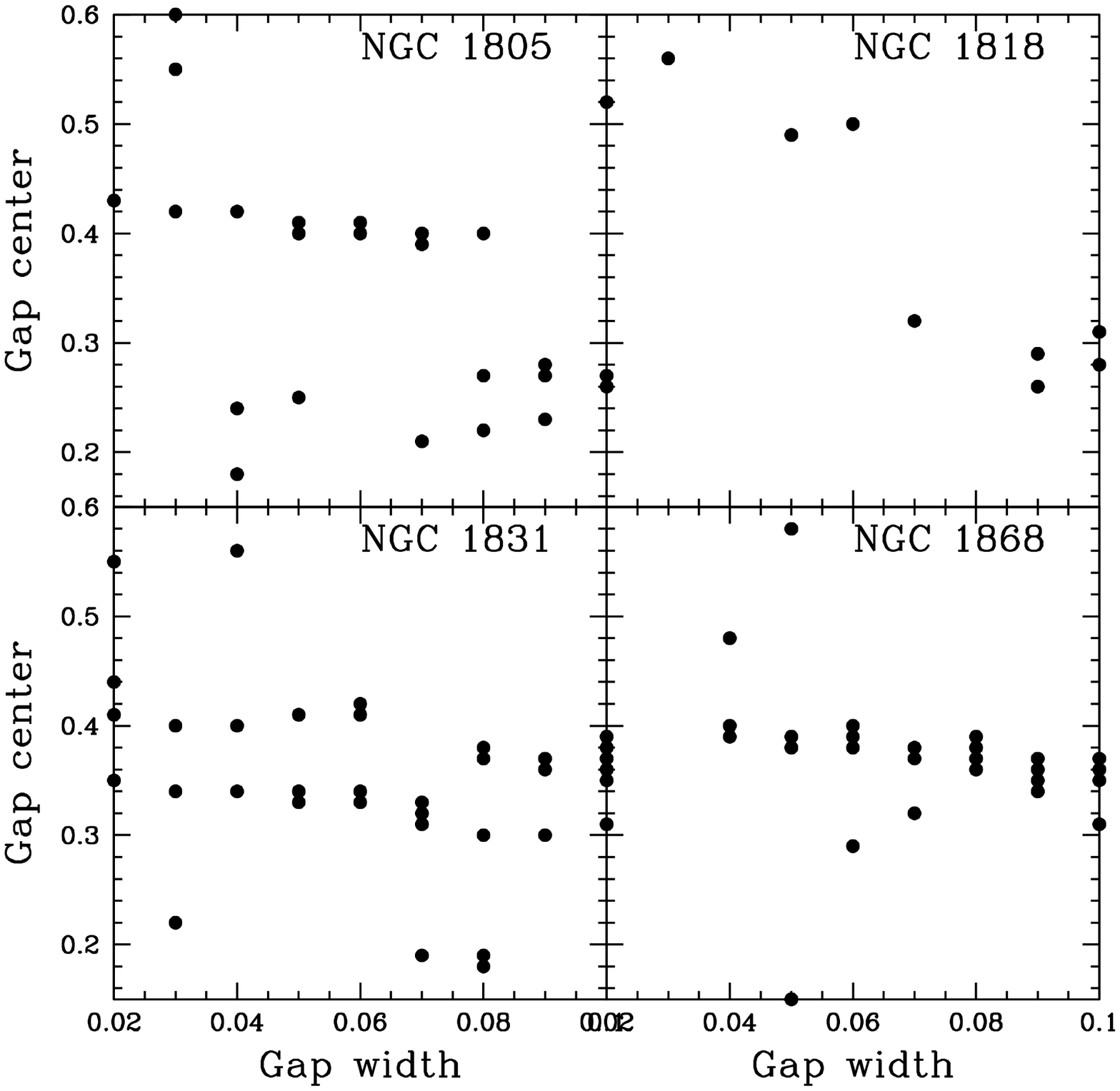,height=10cm,width=9.5cm,angle=0}}
\end{center}
\caption{Panel {\it a}: Gap center plotted against gap width for all high confidence gap candidates found in NGC 1805, according to the $\chi^2$ statistics defined by Rachford \& Canterna (2000). Panel {\it b}: the same as in panel {\it a}, but for NGC 1818. Panel {\it c}: the same as in panel {\it a}, but for NGC 1831. Panel {\it d}: the same as in panel {\it a}, but for NGC 1868.}
\end{figure*}

There is a common pattern in the distribution of these 5\% reddest stars
in the 7 LMC fields. In all cases, the majority of them 
are within the range $20 < m_{555} < 22$. This can be also
seen by the frequency distribution of these stars as a function of
$m_{555}$, which is shown in Figure~7. In all fields,
large $n_{sig}$ stars show a distinct peak in their magnitude histogram
when compared to the distribution of normal MS stars. This peak is
usually broad and covers the $m_{555}$ range mentioned above.
Examples are the fields close to NGC 1805, NGC 1818 
and NGC 1868. NGC 1805 and NGC 1818 are close to star formation regions and
their neighbouring field includes stars of different magnitudes (masses)
leaving the MS towards the RGB. As for the field close to NGC 1868, a 2 Gyr
MSTO is seen in its CMD (Castro et al 2001), contributing with
stars brighter than $m_{555} = 20.5$.

In some cases, on the other hand, high $n_{sig}$ field stars 
show a narrow and very tall peak. This is the case
of the field stars in the 
neighbourhood of NGC 1831, which display a distinct
peak at $m_{555} \simeq 21$; this is 
consistent with the position of the 
$\simeq 4~Gyrs$ turn-off identified in this field by Castro et al (2001). 
Other examples of high peaks in the distribution of detached field stars
include the fields close to Hodge 14 and NGC 2209. These peaks
are at least $0.5$ mag fainter than that of NGC 1831 and
are close to the basis of the SGB of the old LMC field populations.
Inspection of the stars contributing to these peaks in fact confirms that
they are dominated by SGB stars which were not successfully
removed using the $5~\sigma$ colour cut-off. 

Figures 8 and 9 show, respectively, the CMD and $m_{555}$ histogram
for the stars in the on-cluster, background cleaned NGC 1868 
data whose $n_{sig}$ values are at the 95\% position or beyond. 
These figures follow the
same conventions as Figures 6 and 7. A total
of 201 stars contribute to Figures 8 and 9. 
The larger
number of stars allowed a magnitude binning in Figure 9 narrower than in
Figure 7. Notice that the
on-cluster histogram of high $n_{sig}$ stars as a function
of magnitude now shows a large and narrow 
peak at a bright magnitude, $m_{555} \simeq 19.5$,
which is close to the dominant NGC 1868 MSTO.
A second peak is seen in the range $20.5 < m_{555} < 22$, similar
to those found in the field CMDs. This latter peak
is well centered where the candidate secondary MSTO was found.
Apart from the bright peak, however, the distribution of high $n_{sig}$
stars as a function of $m_{555}$
in the cluster data is not markedly different from those of field stars.

Thus, we conclude that a field origin for the second turn-off
in NGC 1868 may not be ruled out. However, 
the differences in shape between this feature
and those known to be associated with field LMC stars, plus the fact that
the NGC 1868 CMD had previously been removed of
field stars, argue against this interpretation.
Furthermore, the suspected secondary turn-off in NGC 1868 
is detached enough from the MS that the two reddest stars contributing
to it actually are located beyond the $+5~\sigma$ line shown in Figure 8
(see Figure~2), and therefore do not contribute to the histogram in Figure 9.

\subsection {Selective unresolved binarism and Am stars}

As mentioned in \S 3, there are other causes, besides a MS turn-off,
that could lead to a set of stars being significantly detached from
the MS towards redder colours. 

Unresolved binarism is one such possibility. 
The presence of a secondary star, whose flux is added to the primary,
will shift the position of the system towards brighter magnitudes in a CMD, 
relative to the
position of the primary. If both are MS stars, the secondary will be
redder, thus also affecting the colour of the system. Unresolved 
binaries are certainly present in any CMD and should be distributed
all over the MS and among evolved stars. Their main effect is to
make the MS broader with a red tail in the colour distribution.
The feature seen in Figures 1 and 2, at $m_{555} \simeq 21$, could
be explained only if there is a specific increase in the fraction of 
cluster binaries at
an absolute magnitude $M_{555} \simeq 2.5$.

The apparent second MSTO occurs at the
luminosity typical of peculiar A stars (Ap, Am, etc).
Due to their chemical anomalies, Am stars
in particular tend to be redder than normal A stars by about 0.05 mag as a
result of increased line blanketing. 
This effect alone is too small to account for the feature in question but 
certainly goes in the right direction. As many
Am stars are also known to be binaries (Carquillat et al 2001, Debernardi et 
al 2001), a combination of blanketing
and enhanced binarism might accommodate the observed feature in the CMD of
NGC 1868. However, more stringent constraints on the frequency
of Am stars and on their binary fraction must be placed in order to test
this possibility. Another problem with this interpretation is that
global binary fraction estimates tend to be smaller in rich clusters than 
in the general field (Mayor et al 1996, Cot\'e et al 1996, Elson et al
1998).

\subsection {The B\"ohm-Vitense gap}

The onset of a convective envelope is known to occur for MS stars of
spectral type A or later. This change in energy 
transport is expected to occur abruptly for stars with 
$T_{eff} \simeq 7500~K$, or $(B-V) \simeq 0.22$ (B\"ohm-Vitense 1958). 
As a result, stars cooler than this critical effective temperature would 
become redder than the slightly hotter ones with fully radiative envelopes
(B\"ohm-Vitense 1970). The resulting discontinuity in the CMD is the so-called 
B\"ohm-Vitense gap and is thought to have an amplitude of $\Delta (B-V) 
\simeq 0.10$. Recent surveys combining precise photometry and distance
measurements from astrometry have shown that a gap does
exist and is located not too far from where expected by theory. But 
the evidence is often weak or subject to selection effects.
Newberg \& Yanny (1998) studied bright and nearby field stars with available
photometry plus distances from Hipparcos and found a gap at 
$0.2 < (B-V) < 0.3$ 
in the CMD of field stars of luminosity class V only. In other words,
the gap is masked out by evolved stars contaminating the apparent
MS in their CMD.
Rachford \& Caterna (2000) detected a gap at $(B-V) \simeq 0.35$
in most nearby open clusters they studied, but they argue that this gap
may be unrelated to the onset of envelope convection. 

We have carried out the same gap detection analysis as Rachford
\& Caterna (2000).
We computed $\chi^2$ values as defined by those authors for
gap candidates within the $0.15 < m_{555} - m_{814} < 0.60$ colour range
and with widths varying from 0.02 to 0.10 mag. 
Figure 10 shows the $m_{555} - m_{814}$ frequency distribution, 
not only for NGC 1868,
but also for 3 other rich LMC clusters in our sample,
each one with more than 5000 stars in their CMD (Santiago et al 2001). 
The clusters
are paired according to age, NGC 1805 and NGC 1818 being younger ($\tau
< 10^{8}$ yrs, Johnson et al 2001) than NGC 1831 and NGC 1868. 
The colour distributions for the older clusters is slightly
flatter than those for the younger ones. Apparent
gaps are seen at $m_{555} - m_{814} \simeq 0.40$ (in all
cases) and at $m_{555} - m_{814} \simeq 0.52$ (except 
NGC 1831). A broader gap is also present at
$m_{555} - m_{814} \simeq 0.25-0.30$ in NGC 1805 and NGC 1818 and possibly in
NGC 1831 and NGC 1868.

Figure 11 shows the gap center plotted against gap width for all our gap 
candidates with $> 90\%$ confidence ($\chi^2 > 2.8$). 
Each panel corresponds to
one of the 4 rich clusters discussed. The gaps at $m_{555} - m_{814} 
\simeq 0.4$ are significant in all clusters except NGC 1818.
At $m_{555} - m_{814} \simeq 0.52$, only narrow and fairly
unconspicuous gaps are confirmed.
The bluer gap, at $m_{555} - m_{814} \simeq 0.25$, takes different widths
and positions, being more significant in NGC 1805 and NGC 1831.
Hence, the main feature, clearly present in 3 rich clusters, is
the gap at $0.35 < m_{555} - m_{814} < 0.40$. 
Notice that this colour coincides closely with that of the candidate MSTO, 
adding support to the statistical reality of this CMD feature. These
gap limits correspond approximately to $0.23 < B-V < 0.27$,
which is close to the expected position of the B\"ohm-Vitense gap
and somewhat displaced from the gap position favoured 
by Rachford \& Caterna (2000).

The presence of colour gaps in most of our rich LMC clusters calls for a
more common mechanism than merger events. Our gap detection
results, therefore, favour the possibility that the B\"ohm-Vitense gap
was detected for the first time in LMC clusters.
On the other hand, our gap colour is inconsistent with the one found
by de Bruijne et al (2000, 2001) in their very accurate Hyades data. 
In fact, the 
very narrow CMDs by these authors provide the strongest limits on the
gap position and associated colour discontinuity, which seems to
be of $\Delta (B-V) \simeq 0.05$. This is significantly smaller 
than the colour difference between MS stars and the secondary MSTO found in 
our NGC 1868 data.

\section{Conclusions}

In this paper we have shown possible
evidence for a second main sequence
turn-off in a deep WFPC2/HST colour-magnitude diagram of NGC 1868.
This feature is clearly visible in the cluster CMD, at 
$m_{555} \simeq 21$ and $m_{555} - m_{814} \simeq 0.4$,
especially after contaminating field stars are removed.
The presence of these stars, as well as of a residual SGB/RGB brighter
than $m_{555} = 21$,
is evidence for a secondary cluster population. 
In fact, previous ground-based data also show a hint of this feature:
the distribution of stars seen redwards of the
main-sequence in the NGC 1868 CMD by Corsi et al. (1994), despite the
large scatter, has an excess of stars at the same position.

Assuming that the feature is really a MSTO associated with NGC 1868,
isochrone fits yield an age of $2.5-3.5~Gyrs$  and a metallicity of
$-0.7 < [Fe/H] < -0.4$ for this subpopulation. 
Using star counts in CMD regions where
we expect only one cluster subpopulation to be present, we estimate the mass 
ratio of the older subpopulation relative to the younger to be in the
range $0.05$ \ltsima $M_{old} / M_{young}$ \ltsima $0.12$. 
This range of values incorporates uncertainties in mass function
slope, in age and metallicity of the secondary population, as well as
uncertainties in the
statistical removal of contaminating field LMC stars.

Even though the candidate second turn-off remains untouched after field
stars are statistically subtracted from the CMD of NGC 1868, the possibility
that this feature is associated with a field star population was 
investigated in detail and cannot be completely ruled out. CMDs of 
field stars from several different positions in the LMC show an
excess of stars located in loci close to that of the NGC 1868 candidate
turn-off. This excess is measured relative to the expected number 
of stars scattered to these loci due to photometric
errors. Most of the features in the field CMDs, however, are broad 
and visually less conspicuous than the apparent secondary 
turn-off in NGC 1868, thus being more consistent with periods of enhanced
field star formation lasting for longer than $1~Gyr$. Another argument
against a field origin for the candidate second turn-off is that its stars
are concentrated towards the cluster centre.

Other possible explanations for the CMD feature 
have been explored. They
include CMD spread due to unresolved binaries, which could mimic a
MS turn-off if binarism is enhanced within a narrow range of primary
star masses. It is interesting to notice that the mass range in question 
is close to that of Am stars, which are redder than normal A stars
and for which a larger than usual binary
fraction may exist. Estimates of binary fraction within globular clusters,
however, yield smaller values than in the field. Besides, the
binary fraction among Am stars is not yet well constrained. It is thus 
unlikely that the CMD feature discussed here is due to Am stars or unresolved
binaries.

A final and exciting alternative would be that 
the B\"ohm-Vitense gap was for the first
time detected in a stellar population outside the Galaxy. The
fact that gaps were found in the colour distribution of two other LMC
clusters besides NGC 1868, all of them at $m_{555} - m_{814} \simeq 0.4$ 
($(B-V) \simeq 0.25$), calls for a more common mechanism than
mergers. In fact, a merger of the type suggested by the presence of
a second and much fainter MSTO as discussed
in this paper should be rare (Vallenari et al 1998, Dieball \& Grebel 2000). 
Most model predictions favour merging
units that are nearly coeval and usually formed through encounters
within the same giant molecular 
cloud, either before or after they are fully formed clusters 
(Fujimoto \& Kumai 1997, Efremov \& Elmegreen 1998, Leon et al 1999).

On the other hand, it
is unclear to what extent the B\"ohm-Vitense gap may mimic a 
main sequence turn-off.
The recent and precise CMD data on the Hyades by de Bruijne et al (2000, 2001)
do not reveal strong turn-off like features associated with the
B\"ohm-Vitense gap candidates found by those authors; the colour discontinuity
in their CMD is of smaller amplitude than necessary to account for
the NGC 1868 feature studied here. Besides, the gap 
associated with our NGC 1868 CMD feature is bluer than the gaps 
found by de Bruijne et al. 

We should point out that the nature of the stars that make up the candidate
MSTO in NGC 1868, which currently is uncertain and accountable for by at
least two astrophysically interesting 
alternatives (merging and B\"ohm-Vitense gap),
may be established by spectroscopic classification,
a task likely to demand large, 8m class, telescopes.

\section*{Acknowledgments}
We thank Gerry Gilmore, Steinn Sigurdsson, Horacio Dottori and Eduardo Bica
for useful discussions. This work was partially supported
by CNPq and PRONEX/FINEP 76.97.1003.00.

\end{document}